# RESEARCH ON DATA INTEGRATION OF OVERSEAS DISCRETE ARCHIVES FROM THE PERSPECTIVE OF DIGITAL HUMANITIES

Rina Su[1], Yumeng Li[2], Xin Yang[3], Xin Yin[4] and Tao Chen[3]

[1]Sun Yat-sen University Library, Guangzhou 510205
[2]School of Journalism and Communication, Jiangxi Normal University, Jiangxi 330022
[3]School of Information Management, Sun Yat-sen University, Guangzhou 510205
[4]School of Information Resource Management,
Renming University of China, Beijing 100872
Corresponding author:chent283@mail.sysu.edu.cn

## ABSTRACT

*The digitization of displaced archives is of great historical and cultural significance. Through theconstruction of digital humanistic platforms represented by MISS Platform, and the comprehensiveapplication of IIIF technology, knowledge graph technology, ontology technology, and other popularinformation technologies. We can find that the digital framework of displaced archives built through theMISS platform can promote the establishment of a standardized cooperation and dialogue mechanismbetween the archives' authorities and other government departments. At the same time, it can embed theworks of archives in the construction of digital government and the economy, promote the exploration of theintegration of archives management, data management, and information resource management, andultimately promote the construction of a digital society. By fostering a new partnership between archivesdepartments and enterprises, think tanks, research institutes, and industry associations, the role of multiplesocial subjects in the modernization process of the archives governance system and governance capacity willbe brought into play. The National Archives Administration has launched a special operation to recoverscattered archives overseas, drawing up a list and a recovery action plan for archives lost to overseasinstitutions and individuals due to war and other reasons. Through the National Archives Administration, theState Administration of Cultural Heritage, the Ministry of Foreign Affairs, the Supreme People's Court, theSupreme People's Procuratorate, and the Ministry of Justice, specific recovery work is carried out bystudying and working on international laws.*

## 1. INTRODUCTION

According to historical reasons, many precious archives of our country are scattered overseas, including many rare historical archives. It is estimated that the number exceeds millions of volumes, mainly distributed in Japan, Southeast Asia, Europe, and North America. The historical archives stored overseas are an important witness to the splendid Chinese culture. It is an indispensable part of works for clarifying the development of Chinese civilization. In recent years, the international archival community and many countries and regions have increased their attention and support for the collection, collation, publication, and digitization of overseas displaced archives. The International Council on Archives has also devoted decades of effort. The emergence of digital humanities as a new interdisciplinary research paradigm has provided many transformative thoughts and methods for the development of archival science and the full utilization of archives themselves. With the help of the multi-dimensional image intelligence system (MISS platform), this study hopes to provide a brand new research model and resource utilization ecology for these precious archives. Thus, it improves the use-value of overseas displaced archives, helps archives displaced abroad "return home" in a special way, promotes cultural cohesion and centripetal force, strengthens the cultural foundation, and promotes cultural confidence construction.

## 2. STATEMENT OF PROBLEM

Overseas displaced archives are spread in various forms in a wide range. Their preservation state is uneven, and many of those that have historical significance are handwritten. There are also different characteristics of handwriting. Displaced Archivers are concerned with the integrity of a country's archival documentation and the continuity of its historical record.[1]Archives are the original records formed by individuals, groups and institutions in the course of their social activities, and they not only carry the memories of individuals and groups, but are also an integral part of the common memory of society. Displaced archivers, for those who formed them, are important documents and records of their social activities, documenting them in their original form. For society, displaced archivers are an important part of society's cultural heritage and play an important role in the maintenance of emotions, cultural transmission and the transmission of civilisation, especially those archives that have unique historical significance. Archives are interconnected, orderly wholes made up of single copies. Part of a displaced archive exists alone as a slice of memory, severing its organic historical links with other archives in terms of source, content, form and time, and its value is not obvious at that time, unable to reveal the whole picture of social and historical activities, and the uneven geographical and institutional distribution leads to a fragmentation in its value. It is only when it is integrated into the whole that it can be valued as such, achieving an overall function that is greater than the sum of its parts. In the case of the Dunhuang Mogao Caves, the caves contain scriptures, ancient texts and documents that document Chinese politics, economics, culture, military divisions, philosophy, religion, literature and art from the 4th to the 11th centuries, and the loss of the Dunhuang archives has caused a discontinuity in Chinese cultural memory. The problem of displaced archivers is widespread in several countries and regions of the world[2].The issue of the return of overseas displaced archives is receiving attention from the international archival community and national archival departments, especially in countries and regions where archives are seriously lost. However, under the current international law recovery mechanism, the physical return of displaced archivers is subject to statutes of limitations, ownership determination, definition of litigation subjects, and legal differences between countries, and digital return is becoming a new model for the return of displaced archivers [3].The value of archives lies in their use, and digital regeneration strikes a reasonable balance between the return of displaced archival entities and the international use of archival resources. Through the orderly organisation of archival resource images, the MISS platform explores a feasible solution for the digital regeneration of displaced archives.

The multi-dimensional image intelligence system (MISS platform) developed by the author's team provides a suitable solution path to realize the "return" and sharing of these archives within the limited scope of time and space. The MISS platform (http://miss.newwenke.com/sas/) can realize the retrieval, indexing, management, and appreciation of image archives, as well as the programming of the background language SPARQL Editor. It's worth noting that this platform takes "image" as the processing object, so the initial sorting of the archives is crucial, which needs us to store them in the format of "image." Based on the above application principles, this research intends to combine popular digital humanity technology such as Deep Learning, Knowledge Map, and Linked Data in the IIIF (International Image Interoperability Framework) to carry out research on the overseas displaced archives.

## 3. RESEARCH OBJECTIVES

Using MISS to realize the digital application of overseas displaced archives inspires the digital use of archives, especially as a platform for archives sharing.

The technical scheme and research ideas adopted in this study have a certain foresight and have been successfully utilized in Digital Humanity Project like "The newspapers within the Republic of China Era, and cultural heritage". As for archival resources, the purpose of this study is to carry out the following research:

1) Based on understanding and mastering the current situations and the spreading history of overseas displaced archives, digitize archives of a specific region or a group with special significance as the research object, providing a reference to carry out mass photocopying, digital return, and utilization of overseas displaced archives.

2) The structure of knowledge related to organizing displaced archives is carried out through Ontology and Knowledge Graph using popular Digital Humanity technology; the semantic annotation and deep organization of image archive content are combined with the IIIF framework and Web Annotation model.

3) Using the MISS platform, based on semantic annotation of image archives, relate and combine knowledge (people, place, time, event) related to different data sets (sources), eliminate information silos between different archival resources, to realize the serialization of archives across time and space.

## 4. RESEARCH DESIGN AND METHODOLOGY

The establishment of Miss platform is the key path to realize the integration of overseas displaced archival resources and the interconnection of different forms of archives in different regions. The design idea of Miss platform even breaks the traditional way of resource organization and reorganizes the discrete data stored in different regions from the perspective of knowledge meta in order to realize the final interconnection of resources. If we summarize the traditional way of organizing archival knowledge, it can be roughly divided into three stages. The first is the form of archival records labeled according to international archival cataloguing rules. The second is the existing form of electronic archival records, which is applicable to the form of machine readable records, and this can be regarded as the modern form of records. The third is also the current attempt to catalog archives in the form of ontology records, which has been discussed a lot in the archival community, and this approach can be regarded as the future form of records.

From the point of view of displaced archives themselves, the background of their creation is deep and the history of their survival is long. War, armed occupation, internal conflict, regime change, federal division, colonial activities, espionage, natural disasters, unintentional carrying, and foreign backups may lead to the preservation of archives in all their countries [4].The complexity of the reasons for the creation of displaced archives determines the longevity of their existence, which almost accompanies the creation of archives themselves. The complexity of the reasons for the creation of discrete archives determines their longevity, which almost accompanies the creation of archives themselves. In addition to their evidential, evidentiary, and intelligence values, archives also have social memory values and identity functions in the context of cultural heritage. The complexity of displaced archives has been aggravated by the above-mentioned reasons, and the different laws and regulations governing the ownership of archives in different countries have led to the lack of an international solution and unified consensus on the return of displaced archives for a long time. Since the second half of the 20th century, academics and the industry have successively proposed different archives on the treatment of displaced archives from the perspectives of cultural heritage, national sovereignty, and archival science, such as territorial origin, sovereignty traceability, functional relevance, and common heritage [5]. However, after more than half a century of efforts by various stakeholders, including archivists, cultural institutions, international organizations, and national governments, disagreements on the return of displaced archives still exist.

The multifaceted role of archives, as war intelligence, government documents, legal evidence and, also, as cultural heritage and social records, has led to different demands from various stakeholders on the issue of displaced archives. From an archival researcher's perspective, the return of displaced records conforms to the "principle of provenance," where records within the same entire collection cannot be dispersed, but this principle is also challenged in the case of displaced records. Especially in the colonial historical background, such as the United Kingdom, the Netherlands established in the colonies of the trade company files, Guangdong Customs archives. From the perspective of state and society, archives are direct records of individual, social and national social activities. The generation process of archives is the generation process of social memory and cultural heritage. Displaced archives should be returned to all countries at the time of their generation. From the perspective of national laws and regulations, there are often differences in the provisions of displaced archives in different legislative systems, such as the German archives seized by the United States during World War II. German law affirms German ownership of the archives, but in American law, the ownership of the archives has been transferred to the United States government [6].The differences in the demands of all stakeholders have led to difficulties in the return and processing of displaced archives. Meanwhile, in the actual process of the return of displaced archives, the return of displaced archives entities is also mixed with the game of interests between countries. Due to the unique value of the archives, all countries regard overseas displaced archives as part of their own cultural heritage system and refuse to return them easily. Taking the encyclopedia ' Yongle Ceremony ' compiled in the Ming Dynasty of China as an example. Due to war, fire, private theft and other reasons, it is scattered in the hands of more than 30 collection agencies in 8 countries and regions. More than 420 copies of the ' Yongle Ceremony ' have been handed down, and China 's domestic collection agencies only have more than 200 copies. In 2010, French President Sarkozy announced to lease 297 volumes of Korean royal documents from the French National Library to South Korea. The ownership is still owned by France, and South Korea can renew them indefinitely. To sum up, it is a complex and arduous task to completely clarify the ownership of displaced archives.

The complex situation of the 20th century has gone away up to now, and international archival diplomacy has become the forerunner of political and economic exchanges between countries. For example, the 70th anniversary of the end of World War II, the National Archives of Australia (NAA) proposed to return the archives of Japanese companies seized during World War II to the National Archives of Japan (Naj) in order to improve exchanges between the two Australian governments in 2015 [8]. Friendship and cooperation between countries has become an important means of solving the problem of displaced archives today. The "common heritage" solution is becoming a viable option for discrete archives, even though in this case the disposition of displaced archives often faces the complexities of reconciling the existing principles [9], especially the return of displaced archival entities. As digital technology continues to mature, "technology sharing" is becoming a "common ground" solution for the repatriation of displaced archival entities, which avoids the discussion of discrete archival entities and their ownership and provides an alternative path for their data integration.

In this paper, we build a multidimensional image wisdom system based on LIBRA (Linked Data, IIIF, Big Data, RDF, AI) technology system according to the above-mentioned displaced archival characteristics. Through the ontology model, the structured archival data and images in different regions, different time zones, and different archival institutions are annotated in multiple layers (metadata layer, image content layer, and semantic association layer), and the annotated contents are stored in the graph database in the form of RDF triples to achieve a unified representation of cross-format archival resources. With the International Image Interoperability Framework（IIIF）, MISS platform will realize the standardized publication of archival resource images, realize the reorganization and reuse of archival resource images based on IIIF APIs, and realize the digital regression of displaced archives.

In the process of reuse and reorganization, MISS platform does not download the original images from archival institutions, but ensures the intellectual property rights of image owners through Imge API calls (Image URI), providing a secure technical solution for digital dissemination and reuse of discrete archival resources. By publishing archival metadata and images on the MISS platform, the information silos between cross-resources are opened, and the sharing and interoperability of archival resources (data and images) across institutional resources is realized. People, text, seals, etc. in archival image resources can be regarded as objects, and AI-based target recognition technology can realize the annotation of image contents and dynamic text recognition of archival images by accessing external OCR interfaces. In specific applications, through Linked Data technology, MISS uses object attributes in the ontology model as a bridge to connect resources, correlates the annotated contents of archival resources with external knowledge bases. MISS uses the object attributes in the ontology model as a bridge to connect resources through Linked Data technology in the application. Moreover, the platform also associates the annotated contents of archival resources with external knowledge bases and presents the associated knowledge in the form of knowledge graphs to achieve data fusion and semantic discovery across resources and enhance the knowledge abundance of archival resources on the basis of digital reorganization in order to show their values more comprehensively.

### 4.1. International Image Interoperability Framework，IIIF

International Image Interoperability Framework is a set of standards that define the interoperability framework for digital libraries through a standard set of application programming interfaces (APIs). It provides a unified way to describe, distribute, and access images on the Web. In June 2015, under the guidance of IIIF, the British Library and 29 non-profit institutions such as Oxford University Library and Harvard University carried out image resource storage to ensure the interoperability and accessibility of global image resource storage. Use images as a carrier to promote the unified display and use of online resources such as books, maps, scrolls, manuscripts, music, and literature. The use of IIIF enables image resource storage institutions to break through the limitations of their resources and fully realize the interoperability of image resources with other collection institutions, which greatly improves the research ability of the institutions in the network data environment. Once IIIF was proposed, it quickly became a research hotspot in the field of GLAM (Art Gallery, Library, Archives, and Museum). At present, most major international cultural heritage research institutions have joined the IIIF framework.

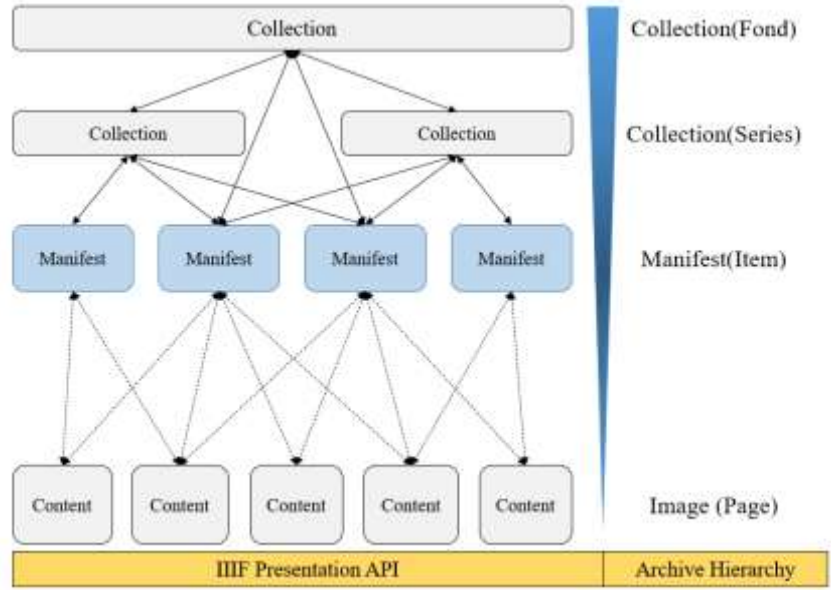

Fig. 1 Organization structure of archival thematic research knowledge

IIIF, as it were, provides an unprecedented new method. It is a set of standards-defined digital library interoperability frameworks that provide a unified method for describing, distributing, and accessing images on the Web through a standard set of application programming interfaces (APIs). This method uses a standardized image request format to share the digital content of images and improves the ability of online research of image resources. Developed through the joint efforts of several institutions, IIIF has quickly been adopted by the wider cultural heritage sector and is receiving more attention in digital humanities construction and research. Dispersal of archival digital resources is where the IIIF framework is put into use. Figure 1 shows the organizational structure diagram of archival image resources using IIIF, and "interaction" becomes the core idea of the whole architecture. 1. "Image interaction": the lowest level is the archival image resources collected by various institutions. These resources are organized into Manifest files through Canvas in the IIIF framework. The archival images contained in the Manifest file are interactive, that is, the archival images of different institutions can be shared and interacted with each other and cross-organized into the Manifest file of archives. 2. "Archive Interaction": Manifest files organized according to different archive images can also interact with each other, thus organizing them into "Archive Collection." 3. "Topic Interaction": Different research topics can also interact with each other and be reorganized into a set of topics at a higher level.

### 4.2. Linked Data

Linked data has become a key technology in digital humanities research in recent years, especially the integration of heterogeneous resources from multiple sources in interdisciplinary research. Notes that associated data and data association are not synonymous. All associated data can be regarded as data association. Linked Data is a lightweight implementation of the Semantic Web. It is not new data, but a new form of data presentation. Linked data is generally considered only if it conforms to the four principles of Linked Data outlined by Tim Berners-Lee in 2006.

1. Use URIs as names for things.
2. Use HTTP URIs so that people can look up those names.
3. Provide useful information about the thing when its URI is dereferenced, using the standards (RDF, SPARQL).

4. Include links to other related URIs in the exposed data to improve the discovery of other related information on the Web.

The main benefits of using Linked Data are:

1. More convenient data access: Your data can be accessed immediately in a machine-processable way through persistent URLs. In terms of sharing data, data from the persistent URI (http://www.mycompany.com/branches/1) is more efficient than in the data warehouse data (for example, the isolation of excel files and even private commercial database server)
2. The schema is more flexible: Your linked data does not follow a particular schema, there are no database tables, just a bunch of declarations, and you can add more statements whenever you want. Or let another data provider supplement the data by adding more statements.
3. More standard query language: Using popular or standardized vocabularies increases data interoperability and allows queries to be performed across multiple data repositories.
4. Data interconnection is more intelligent: the main function of linked data is to link the data with external resources. By linking to your data, you can enrich your data by sharing any other valuable information on the network.

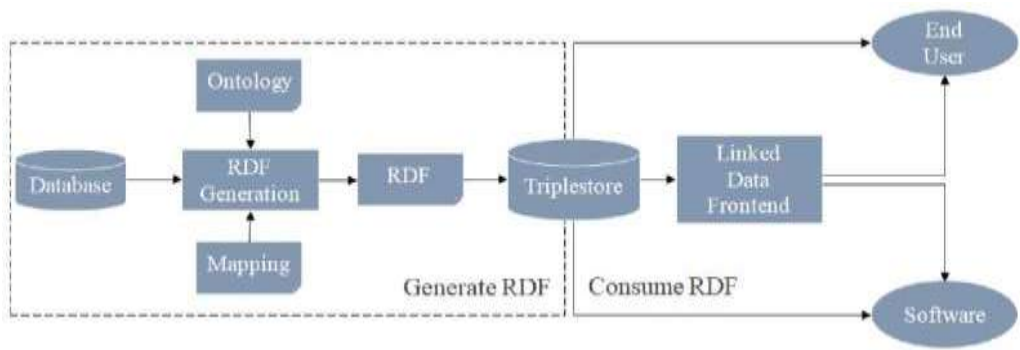

Fig. 2 Process framework of the linked data system

Fig. 2 shows the typical process framework of the linked data system, which is mainly divided into two parts: the generation and consumption of RDF data. Generate

RDF mainly aims at the processing process of RDF data, and RDF conversion is mainly carried out on archive data existing in the Database. During the conversion, the corresponding ontology should be designed, and the program coding should be carried out to map the ontology and archive field information in the Database. The resulting RDF data is recommended to be stored in the Triple Store. Currently, the threshold for structuring data RDF is getting lower and lower, and many mature tools (D2R, Open Refine, and R2RML) can easily convert it. During the Consume RDF phase, RDF data in the Triple Store can be directly called using an interface, such as publishing the data in the Triple Store using the SPARQL Endpoint. Of course, we can also develop some front-end interactive pages with a better interactive experience.

## 4.3. Ontology

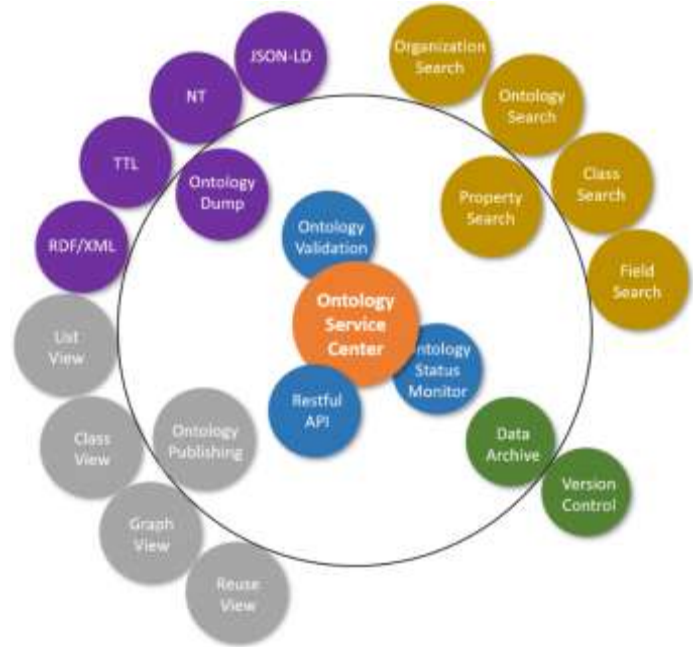

Fig. 3 Functional architecture of Ontology Service Center

The construction of the knowledge graph of overseas displaced archives mainly involves ontology construction, RDF structuring of data, and resource association among different data sources. Firstly, ontology is used to organize the owners, units, semantic-tagged content, and other data of the scattered archives. At present, there are more than 700 native word lists commonly used in LOV (Linked Open Vocabularies). In addition, the Ontology Service Center built by the research group also contains a large number of ontologies related to the digital humanities field. In addition, the Graphic Library in China has released lists of common words in the field of digital humanities, such as people, places, and times.

The ontology service center (OSC) framework, shown in Figure 3, composes of four main components: Ontology Visualization, Property Search, Version Control, Ontology Reuse, Ontology Validation, Ontology Publishing, and Ontology Status Monitor.

1) **Data Dump**.

OSC provides four data dumps, RDF/XML, TTL, RDF/JSON, and NT, which will be generated automatically real-time of ontology (RDF) model that is a set of Statements of this ontology.

2) **Data Publishing/View**.

The system supports four kinds of ontology view methods, Class View (C.V.), List View (L.V.), Graph View (G.V.), and Reuse View (R.V.).
-   Class View. This view uses a tree structure to display the ontology, which makes it easy to understand the hierarchical relationship between classes in the ontology.
-   List View. This view presents the entire vocabulary on one page with an ordered list of classes and properties, followed by more detailed information panels further down the document.

International Journal of Web & Semantic Technology (IJWesT) Vol.14, No.1, January 2023

- Graph View. This view visually displays the ontology with WebVOWL, a web application for the interactive visualization of ontologies. The visualization is automatically generated from the ontology graph.
- Reuse View. This view is achieved through the WebVOWL Editor application, which is designed to serve the skills and needs of domain experts with limited knowledge of ontology modeling.

**3) Data Archive**.

The archived ontology is saved as a file on disk, and the file name will contain the version number of the archive. In other words, only the latest ontology will be stored in the RDF store. The system will provide the results of the comparison between the archived version and the latest one. If necessary, the archived ontology can be rolled back and restored to the previous version in the ontology graph.

**4) Data Search**

OSC enables searching for vocabulary terms (class, object property, data property), term comments, catalogs, namespace prefixes, and contributors. In addition, you can perform ontology state monitoring, ontology validation, and RESTful interface APIs.

**5) Empirical case analysis**

Based on the core technology and framework, we construct a multi-dimensional image intelligent system (MISS) platform. The platform supports online sorting, publishing, reuse, semantic annotation, and other functions of image resources in various formats, and now supports online interaction of large-scale image resources, providing a solid technical fortress for the reuse of cultural heritage.

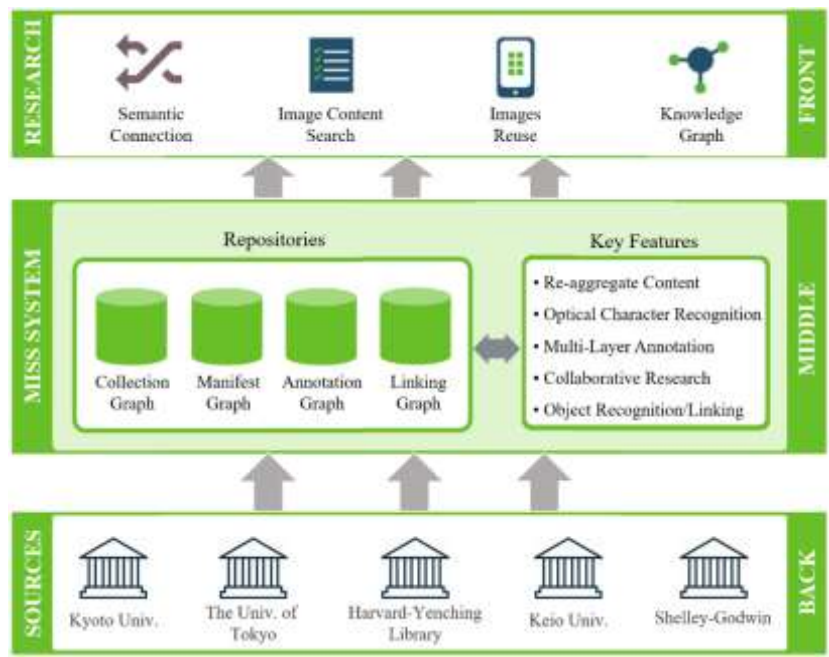

Fig.4 MISS solution for displaced archives

Fig 4 shows the solution for the integration of displaced archives using the MISS platform, which is mainly divided into three layers: front-end application, mid-level integration, and back-end resource import. At the bottom are archival resources from different institutions, which should ideally be published and shared under the standards required by the IIIF framework. The top layer is the specific archival research direction, mainly including Semantic Connection, Image Content Search, Archive Images Reuse, and Knowledge Graph. The middle layer provides the core functions and storage structure for the MISS platform. In the MISS platform, one can perform content re-aggregate, optical character recognition, multi-layer annotation, collaborative research, and object recognition operations. When using the MISS platform to integrate and organize archives, storage will be carried out according to the structure required by the IIIF framework, mainly using Collection Graph, Manifest Graph, Annotation Graph, and Linking Graph.

Next, some innovative ideas for the MISS platform in the utilization of archive resources are described below.

1) **"One-click" lower technical threshold**

   MISS platform provides a "one-click" process for publishing and reusing archive images. Users can not only upload private archive images and generate IIIF resources but also import internet archive manifest resources. Through "One-click", external resources can be reorganized on the platform. In addition, when reusing the archive image, the original image address will be inherited to the new manifest and form an image "gene chain."

2) **Multi-model annotation for "close reading"**

   According to the characteristics of Overseas Displaced Archives image resources, we propose a three-layer annotation model: image layer, object layer, and semantics layer. This model can enrich the content and cluster and link the objects in the images, which is convenient for users to quickly obtain and understand the deep meaning of images.

   The image layer annotation is copied into the archive data, that is, the structured metadata in the archive institution is converted and stored in the form of triples. The object layer annotation is a triplet description for image content ( characters, seals, etc. ) to achieve precise association based on image objects, such as the same Tibetan seal in different images. The semantic layer uses the linked data technology to semantically associate the annotation content of the image layer and the object layer with other linked data sets, and aggregates the knowledge into the MISS platform in the form of knowledge graph.

3) **The symbiosis between images and text**

   OCR separated from images will lack the soul of Chinese characters. Therefore, the concept of real-time OCR is proposed, and third-party APIs are transferred to perform real-time OCR and manual proofreading of the text on the image. The generated OCR text can also be used as a machine learning corpus to improve the OCR accuracy. The concept of real-time OCR reduces the complexity of the conversion process of images and texts and breaks them into parts, which can better assist researchers in humanities.

4) **Infrastructure construction and transmission of Overseas Displaced Archives**

   IIIF includes images from different countries and institutions and forms the "Image of Web." As more and more organizations join IIIF-C, the images become more and more valuable. We believe that the IIIF will become the basic framework for the entire country and even the global archive organizations and further form the ecosystem. In the information technology age, openness and integration can be creative.

Here, taking as an example the collection archives of Keio University Library, Harvard-Yenching Library, Kyoto University Library, and Chester Beatty. Online integration of the resources of these institutions is made in the MISS platform.

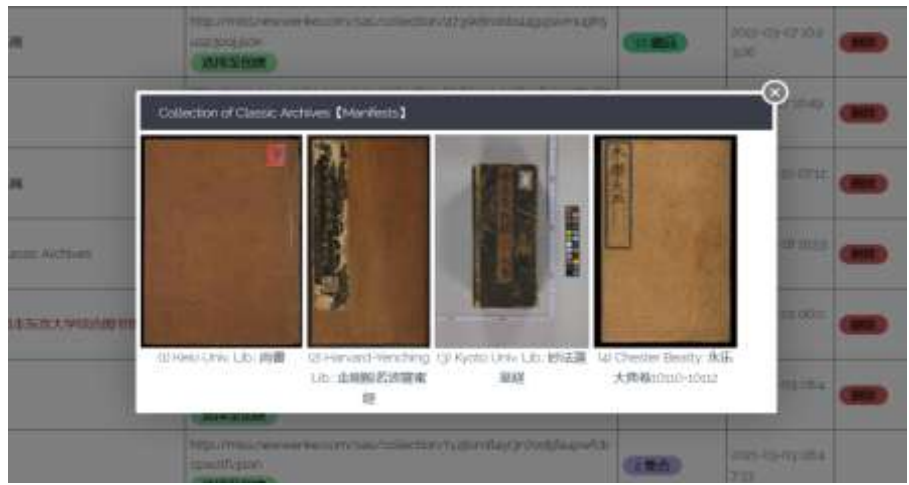

Fig. 5 Collection of Classic Archives Preview

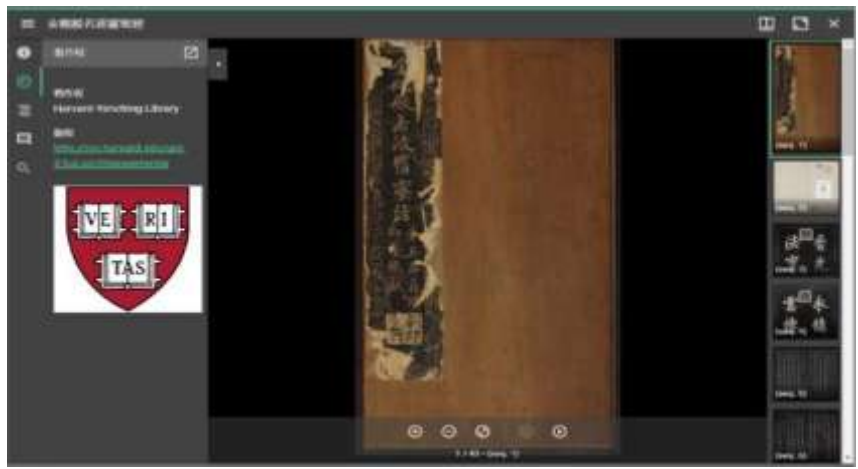

Fig. 6 Full text of the Vajra Prajnaparama Sutra

Firstly, import the Manifest URI address of the archival resources of the four institutions into the MISS platform during integration.

Secondly, after importing, the resources of each institution can be presented independently. That is, a new Manifest URI address can be generated in the MISS platform. Then, select the Manifest of resources of these institutions in MISS to generate a new Collection.

Figure 5 shows the preview screen for the new collection, where you can see the archives from the four sources. Note that the left and right operations in the MISS platform do not download digital images of online files. Figure 6 shows the screenshot of browsing the full text of Harvard-Yenching

International Journal of Web & Semantic Technology (IJWesT) Vol.14, No.1, January 2023

Library archives (Vajra Prajna Paramita Sutra) using Mirador 3.

After importing archive resources into the MISS platform, OCR recognition, semantic annotation and association can be carried out on the platform. Further research on the knowledge graph can even be carried out. In addition to the above functions, displaced archives can also use this platform to achieve cultural discovery of displaced archives. As a separate file or cultural heritage, the value of some displaced archives is limited. By configuring the SPARQL retrieval port of archive resources, the MISS platform can not only realize semantic-based displaced archive retrieval, but also provide cross-knowledge-base retrieval results based on resource similarity by means of cross-knowledge-base federated query, that is, to retrieve the labeled object information (triples) in archive resources in SPARQL Endpoint of different linked datasets. Currently, MISS has been able to aggregate the URIs of the same author from the CBDB, DBpedia, WIKIDATA and other linked data sets to enrich the information about the author and achieve the effect of cultural discovery, so as to create space for the value enhancement of displaced archive data beyond digital reorganization. Similar to the author information, other annotations in archival image resources can also be used for cultural discovery.

## 5. SUMMARY AND PROSPECT

In conclusion, according to the literature research, a large number of all the archival resources in China have been lost in museums and other institutions in Japan, the United Kingdom, the United States, and other countries due to wars. The cost for redeeming them is exorbitant, which is negative for China's national dignity and the dignity and value of the archives themselves. Therefore, the redeeming need to base on the strategic thinking of great power diplomacy. Effectively safeguarding the ownership and right of the recoursing of China's lost overseas archives, the integrity of China's national memory, and the national attribution of archives is imperative. According to the need of building cultural confidence, it is urgent to search for the scattered archives lost overseas due to war and other reasons at this stage, to maintain the dignity and integrity of Chinese culture. According to the promotion of the status of the archival discipline, the promotion of the discourse power of the discipline, and the need for the development of the discipline, China's archival discipline can explore the unique value of the discipline and improve the status of the discipline by carrying out the successful operation of recovering overseas displaced archives. To solve the discussed archives above, we propose four proposals：

First, promote the establishment of a normal cooperating and dialogue mechanism between archives authorities and related government agencies, embed archives work into the overall framework of digital government construction, digital economy development, and digital society construction, and explore the integrated development of archives management, data governance, and information resource management.

Second, foster a new partnership between archives departments and enterprises, think tanks, research institutes, industry associations, and other social organizations, letting multiple social subjects play a role in the modernization process of the archives governance system and governance capacity.

Third, the National Archives Administration led a special operation to recover overseas displaced archives, drawing up a list and a recovery action plan for archives lost to overseas institutions and individuals due to war and other reasons.

Forth, carries out specific recovery work through the study and work of international law by the National Archives Administration, in conjunction with the National Cultural Heritage Administration, the Ministry of Foreign Affairs, the Supreme People's Court, the Supreme People's Procuratorate, and the Ministry of Justice.

ACKNOWLEDGMENTS

This research is granted financial support from the Research Project of Chinese Library Association(2022LSCKYXM-ZZ-ZD002).